# Anisotropic Triangular Lattice Realized in Rhenium Oxychlorides $A_3ReO_5Cl_2$ (A = Ba and Sr)

*Daigorou Hirai[†,\*], Takeshi Yajima[†], Kazuhiro Nawa[‡], Mitsuaki Kawamura[†], and Zenji Hiroi[†,\*]*

[†]Institute for Solid State Physics, University of Tokyo, Kashiwa, Chiba 277-8581, Japan

[‡]Institute of Multidisciplinary Research for Advanced Materials, Tohoku University, 2-1-1 Katahira, Sendai 980-8577, Japan

**ABSTRACT:** We report the synthesis, crystal structure, and magnetic properties of two new quantum antiferromagnets $A_3ReO_5Cl_2$ (A = Sr and Ba). The crystal structure is isostructural with the mineral pinalite $Pb_3WO_5Cl_2$, in which the $Re^{6+}$ ion is square-pyramidally coordinated by five oxide atoms, and forms an anisotropic triangular lattice (ATL) made of $S$ = 1/2 spins. The magnetic interactions $J$ and $J'$ in the ATL are estimated from magnetic susceptibilities to be 19.5 (44.9) and 9.2 (19.3) K, respectively, with $J'/J$ = 0.47 (0.43) for A = Ba (Sr). For each compound, heat capacity at low temperatures shows a large $T$-linear component with no signature of long-range magnetic order above 2 K, which suggests a gapless spin liquid state of one-dimensional character of the $J$ chains in spite of the significantly large $J'$ couplings. This is a consequence of one-dimensionalization by geometrical frustration in the ATL magnet; a similar phenomenon has been observed in two compounds with slightly smaller $J'/J$ values: $Cs_2CuCl_4$ ($J'/J$ = 0.3) and the related compound $Ca_3ReO_5Cl_2$ (0.32). Our findings demonstrate that 5d mixed-anion compounds provide a unique opportunity to explore novel quantum magnetism.

# 1. INTRODUCTION

In most magnetic compounds, a long-range order (LRO) takes place at a low temperature on a scale to match the energy of magnetic interactions between spins. Alternatively, in the presence of strong quantum fluctuations, the LRO is replaced by an exotic electronic state called the quantum spin liquid (QSL). A typical QSL is a Tomonaga-Luttinger liquid (TLL) known as the ground state of a one-dimensional (1D) antiferromagnetic Heisenberg spin-1/2 chain[1,2], in which quantum fluctuations are enormously enhanced owing to the small numbers of neighboring spins and the small spin quantum number. In higher dimensions with more neighbors, however, quantum fluctuations may not be sufficient to maintain such QSL states as the ground states. On the other hand, Anderson proposed that an additional ingredient, that is the geometrical frustration, can enhance fluctuations and stabilize an exotic liquid-like ground state called the resonating valence bond (RVB) state even in two- or three-dimensional (2D or 3D) frustrated systems, such as triangular lattice antiferromagnets[3]. Lured by this intriguing theoretical prediction, enormous studies have been devoted to realizing QSLs in 2D and 3D systems[4].

The ground state of the 2D triangular lattice antiferromagnet, which was initially predicted to host an RVB state[3], is now established both theoretically and experimentally to be a noncollinear LRO with neighboring spins forming 120° angle to each other[5–8], as depicted in Fig. 1a. On the other hand, an anisotropic triangular lattice (ATL) with two kinds of magnetic interactions, $J$ and $J'$, has been focused[9–18]; the ATL model bridges the regular triangular lattice at $J = J'$ and isolated 1D spin chains in the limit of $J' = 0$ (Fig. 1a). The theoretical study shows that a 120° LRO is robust at $J'/J$ larger than 0.8, while, interestingly, a TLL-like gapless QSL state appears in a wide parameter range of $J'/J$ smaller than 0.6–0.7[9–16]. As schematically shown in Fig. 1b, the interchain interactions are effectively canceled owing to the geometrical frustration when an antiferromagnetic correlation develops within the chains upon cooling below $\sim J/k_B$, which makes the ATL magnet into a set of decoupled 1D spin chains. As a consequence of this "one dimensionalization" by frustration[15], a TLL-like QSL state is realized in the apparently 2D lattice with significantly large $J'$ couplings. On the other hand, the ground state realized for the intermediate anisotropy region at $J'/J$ between 0.6 and 0.8 is still in debate: various ground states, such as gapped QSL[10,16], a spiral LRO[17,18], and a dimer-ordered phase[9] are expected to appear in this region; they are nearly degenerate owing to the strong frustration.

On the experimental side, a few model compounds for the ATL have been studied thus far. The most intensively investigated is $Cs_2CuCl_4$ with $J$ = 4.3 K and $J'/J$ = 0.3[19–22], in which inelastic neutron scattering experiments revealed QSL-like spin excitations[19,20] which is interpreted as evidence of the one-dimensionalization by frustration in the ATL[14]. Another well-studied compound is κ-(BEDT-TTF)$_2$Cu$_2$(CN)$_3$, which has $J$ = 250 K[23] with $J'/J$ = 0.3–1[24–27]. For a better understanding of ATL magnets, more compounds in the wide range of $J'/J$ are required.

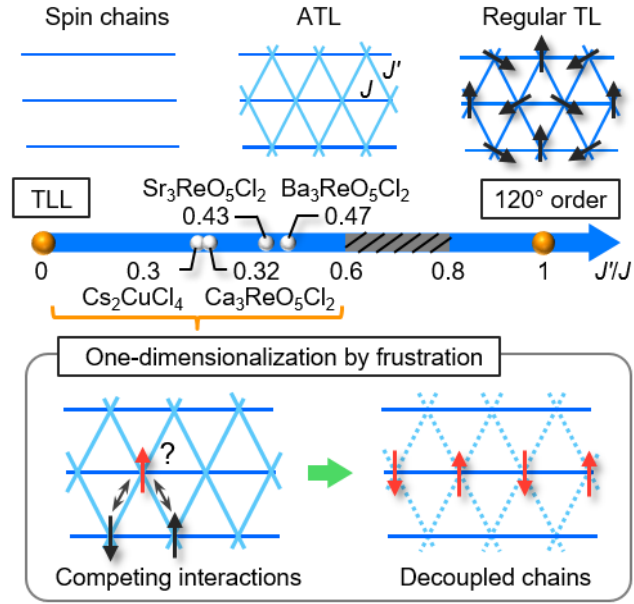

Figure 1. Schematic phase diagram of the Heisenberg spin-1/2 anisotropic triangular lattice (ATL) magnet as a function of the anisotropy in the magnetic interactions $J'/J$. In the limit of $J'/J$ = 0, the ground state is known as a Tomonaga-Luttinger liquid (TLL) for a single spin chain, while the ground state of a regular triangular lattice (TL) magnet at $J'/J$ = 1 is the 120° spin order. The ground states for intermediate $J'/J$ values are a TLL-like QSL for $J'/J$ < 0.6 and a 120° order for $J'/J$ > 0.8, while, for 0.6 < $J'/J$ < 0.8, various ground states, such as gapped QSL[10,16], a spiral LRO[17,18], and a dimer-ordered phase[9] have been proposed. Placed along $J'/J$ axis are the candidate ATL magnets; $Cs_2CuCl_4$[19–22] and $Ca_3ReO_5Cl_2$[28,29] thus far studied, and $Sr_3ReO_5Cl_2$ and $Ba_3ReO_5Cl_2$ presented in this work. The lower panel shows a schematic representation of one-dimensionalization by geometrical frustration. The $J'$ couplings with two neighboring spins on either of the adjacent chains compete with each other and are effectively canceled out

when antiferromagnetic correlations develop in the chain at low temperatures below the order of *J*, leading to decoupled spin chains.

We have focused on mixed-anion compounds[30] with 5d transition metal ions as new material platforms for ATL magnets. Compared with $Cu^{2+}$ ions with the $3d^9$ electronic configuration or $Ti^{3+}$ or $V^{4+}$ ions with the $3d^1$ configuration[31–39], 5d transition metal ions with high oxidation states have been explored less thus far. In addition, a mixed-anion configuration around a transition metal ion often gives a unique crystal structure with anisotropic chemical bondings among magnetic ions. Furthermore, it is advantageous for stabilizing a spin-1/2 state instead of a complex spin–orbit-entangled state expected for 5d electrons by completely quenching the orbital moments.

Recently, we succeeded in synthesizing a new ATL magnet $Ca_3ReO_5Cl_2$ (CROC)[28,29]. As shown in Fig. 2b, the crystal structure consists of the 2D $Ca_3ReO_5$ layers and Cl ions locating between the layers. The $Re^{6+}$ ion with the $5d^1$ electronic configuration, which carries spin-1/2, is surrounded by five oxide ions and one chloride ion, forming a heavily distorted $ReO_5Cl$ octahedron. The crystal field from the surrounding oxide and chloride ions lifts the degeneracy of the 5d levels of Re ions, and stabilizes the $d_{xy}$ orbital at the lowest energy to be singly occupied. Since the orbital angular momentum is quenched by the crystal field, CROC can be regarded as a spin-1/2 Heisenberg quantum magnet; the effect of spin–orbit interactions (SOIs) appears in the reduced *g*-factor of 1.78[29].

The $ReO_5Cl$ octahedra of CROC do not share their anions, keeping the nearest-neighbor distance longer than 5 Å, which results in localized 5d electrons with small electron transfers; the compound is an insulator. The $d_{xy}$ orbitals of $Re^{6+}$ ions lie along the *bc* plane to form an ATL with the anisotropy of $J'/J = 0.32$[29], which is comparable to that of $Cs_2CuCl_4$. The signature of TLL was observed in thermodynamic properties such as a Bonner-Fisher-like magnetic susceptibility and a large *T*-linear magnetic heat capacity at low temperatures[29]. Thus, one-dimensionalization by geometrical frustration really occurs in CROC, as in $Cs_2CuCl_4$.

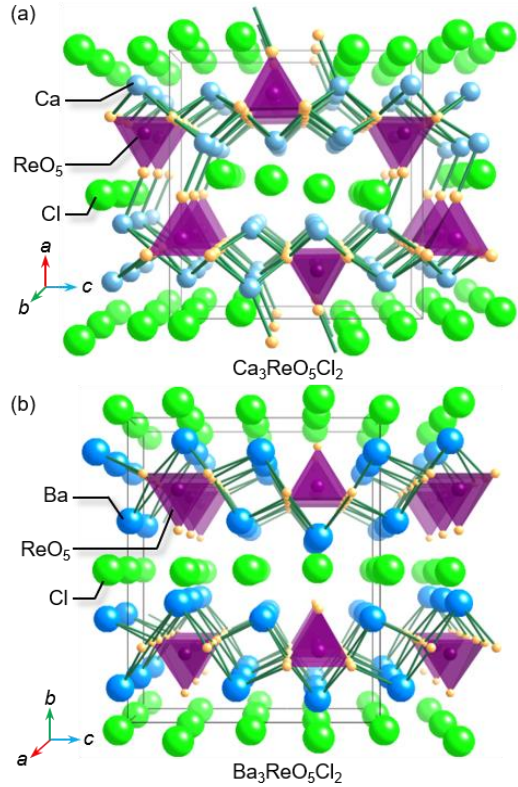

Figure 2. Crystal structures of (a) $Ca_3ReO_5Cl_2$ and (b) $Ba_3ReO_5Cl_2$ viewed perspectively along the [100] and [010] direction, respectively. Either of the compounds consists of an alternate stacking of the $A_3ReO_5$ and Cl layers. A difference comes from the arrangements of the Cl atoms: they form a 2D sheet in $Ba_3ReO_5Cl_2$, while they are segmented into slabs in $Ca_3ReO_5Cl_2$. The difference may be ascribed to the small ionic radius of Ca compared with Ba.

There have been known several compounds related to CROC with the common chemical formula $A_3MO_5X_2$, where A is an alkaline earth ion or $Pb^{2+}$, X is a halogen ion such as $Cl^-$ and $Br^-$, M is a hexavalent cation such as $W^{6+}$ and $Mo^{6+}$[40–43]. For example, $Ca_3WO_5Cl_2$ is isostructural to CROC[40], and $Pb_3WO_5Cl_2$, known as a natural mineral "pinalite," has a similar but slightly different layered structure[41,43], in which the $A_3MO_5$ layers alternate with the flat $X_2$ layers without such a corrugation as observed in CROC (Fig. 2); the difference may come from the difference in the ionic radii of A cations. All the reported $A_3MO_5X_2$ compounds, except CROC, do not contain magnetic ions. We expect new quantum magnets to be synthesized by replacing $W^{6+}$ with magnetically active $Re^{6+}$ with similar ionic radii (0.60 and 0.55 Å in the six coordination, respectively) [44].

Here, we report the synthesis and physical properties of new 5d ATL magnets $Sr_3ReO_5Cl_2$ (SROC) and $Ba_3ReO_5Cl_2$ (BROC), which were discovered in the course of materials search for mixed-anion compounds. These compounds crystallize in a pinalite-type structure comprising ATLs made of $Re^{6+}$ ions. Magnetic susceptibility measurements reveal that SROC and BROC have $J'/J$ values larger than that of CROC (0.47 and 0.43, respectively) and no LROs above 2 K. Instead, a large $T$-linear magnetic heat capacity characteristic for gapless excitations in 1D spin chains is observed for each compound, indicating that one-dimensionalization by frustration occurs to stabilize a TLL-like state even in these large $J'/J$ ranges. Our findings demonstrate that mixed-anion compounds with 5d transition metal ions provide us with a promising platform searching for spin-1/2 quantum magnets.

## 2. EXPERIMENTAL

Polycrystalline samples of $A_3ReO_5Cl_2$ were synthesized by a conventional solid-state reaction. AO and $ACl_2$ (A = Ba, Sr) and $ReO_3$ were mixed at a molar ratio of 2:1:1 in an argon-filled glove box, and the mixed powder was pressed into a pellet. The pellet was wrapped in a gold foil and then sealed in an evacuated quartz tube; the gold wrap was necessary to prevent reaction with the quartz tube. The tube was heated at 750 °C for 24 hours. The sintered pellet was reground, pelletized, and heated again at 800 °C for 48 hours in an evacuated quartz tube. Single crystals of $A_3ReO_5Cl_2$ with a size of approximately 50 μm were grown at 1050 °C by partially melting a stoichiometric mixture in an evacuated quartz tube. Transparent dark orange plate-like crystals grew on the surface of the pellet.

The crystal structures of $A_3ReO_5Cl_2$ were determined by single-crystal X-ray diffraction (XRD) measurements, which were conducted at room temperature using an R-AXIS RAPID IP diffractometer (Rigaku) with monochromated Mo–Kα radiation. The structure was solved by direct methods and refined by full-matrix least-squares methods on $|F^2|$ by using the SHELXL2013 software.

The polycrystalline samples of BROC and SROC used for physical properties measurements were characterized by powder XRD (Rigaku RINT-2500) using Cu–Kα radiation. All the peaks in the XRD patterns are consistent with the structure determined by single-crystal XRD measurements, indicating phase pure polycrystalline samples were synthesized.

Magnetic susceptibility measurements were conducted in a magnetic properties measurement system (MPMS-3, Quantum Design) using approximately 30 mg of pelletized polycrystalline samples fixed on a quartz sample holder with GE varnish (General Electric #7031). Low-temperature heat capacities were measured based on a relaxation technique in a physical properties measurement system (PPMS, Quantum Design).

First-principles calculations were performed based on the density functional theory, using the program package Quantum ESPRESSO[45] which employs plane-waves and pseudopotentials to describe the Kohn-Sham orbitals and the crystalline potential, respectively. The plane-wave cutoff for a wavefunction was set to 60 Ry. Calculations were performed with a GGA-PBE[46] functional using ultrasoft pseudopotentials[47]. We set k-point grids of Brillouin-zone integrations for the charge density and the partial density of states to 10×10×5 and 20×20×10, respectively. Wannier functions were obtained by using a program package Wannier90[48], which computes the maximally localized Wannier orbital.

## 3. RESULTS AND DISCUSSION

### 3.1. Crystal structure.

Our structural analyses using single-crystal XRD data have revealed that BROC and SROC are isostructural to pinalite ($Pb_3WO_5Cl_2$)[42], which crystallizes in an orthorhombic structure with the space group *Cmcm* (No.63); $Ba_3WO_5Cl_2$ also has the same type of structure[49]. The detailed results of the structural analyses are shown in Table SI. The lattice constants obtained are $a$ = 5.79424(18) Å, $b$ = 13.9508(4) Å, $c$ = 11.4414(5) Å for BROC, and $a$ = 5.6492(3) Å, $b$ = 13.1886(6) Å, $c$ = 11.1144(5) Å for SROC. All the lattice constants of BROC are larger than those of SROC because of the larger ionic radius of $Ba^{2+}$ (1.42 Å in the eight-fold coordination) than $Sr^{2+}$(1.26 Å). As is expected from the similar ionic radii of $Re^{6+}$ and $W^{6+}$[44], the lattice constants of BROC is comparable to those reported for the W analog $Ba_3WO_5Cl_2$: $a$ = 5.796(2) Å, $b$ = 13.825(2) Å, and $c$ = 11.469(2) Å[49].

Figure 2b shows the crystal structure of BROC, which has an alternate stacking of $Ba_3ReO_5$ and $Cl_2$ layers along the *b* direction. As compared in Fig. 2, the difference between this structure and that of CROC structure is in the coordination of A cations: the coordination numbers for the two A cations in BROC are 8 and 9, while those in CROC are 5 and 6 because of the smaller ionic radius of $Ca^{2+}$. As a result, Cl ions form a square lattice in BROC (Fig. 2b), while an array of segmented slabs running along the *b* direction in CROC (Fig. 2a).

The $A_3ReO_5$ layer in BROC and SROC is structurally related to α-PbO(litharge)[50], as depicted in Fig. 3. The crystal structure of α-PbO contains square nets of oxide atoms with Pb atoms above or below the center of the oxygen squares in a checkerboard pattern (Fig. 3a). In contrast, there are corrugated square nets of oxide ions in the $A_3ReO_5$ layer with A and Re atoms located above or below the net in a similar manner. As shown in Fig. 3b, three-quarter of Pb atoms in α-PbO are replaced by the A atoms in the $A_3ReO_5$ layer, and the rest one-quarter are replaced by the Re-O units, resulting in the $A_3[ReO]O_4 = A_3ReO_5$ composition. The $Re^{6+}$ ion is surrounded by four oxygen atoms forming a square lattice (O1) and one apical oxygen (O2) into a $ReO_5$ square-pyramid. Note that there is only one crystallographic site for Re, while there are two types of $ReO_5$ square-pyramids: one pointing up and the other pointing down; these $ReO_5$ are connected by inversion symmetry. The $ReO_5$ square-pyramids pointing upward and downward are arranged along the *a* axis, and alternate along the *c* axis. Three neighboring $Re^{6+}$ ions form an isosceles triangle elongated along the *c* axis, and 2D ATL of the $Re^{6+}$ ions is generated in the *ac* plane, as shown in Fig. 3b.

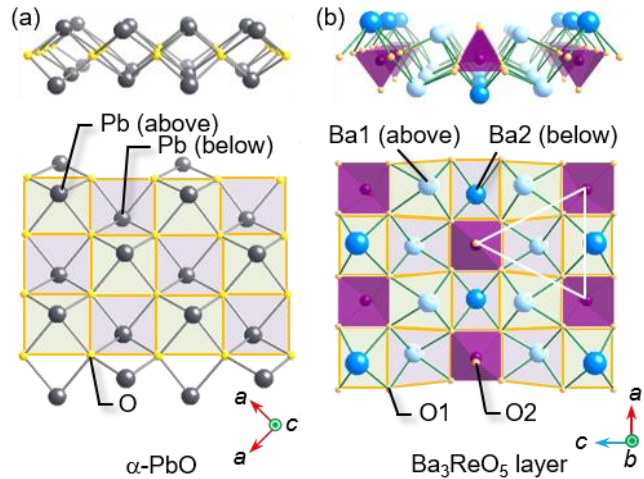

Figure 3. Comparison between (a) the PbO layer in α-PbO and (b) the $Ba_3ReO_5$ layer in BROC. In (a), the yellow squares represent the square net formed by oxide atoms. Pb atoms are located above or below the centers of the $O_4$ squares when viewed along the *c* direction in a checkerboard pattern. In (b), the Ba atoms and Re-O2 units in the $Ba_3ReO_5$ layer are arranged in a similar manner as the Pb atoms in α-PbO. Three neighboring Re ions form an isosceles triangle elongated along the *c* axis. The Re sublattice generates

an anisotropic triangular lattice.

Table 1 Selected bond lengths ($l$) for $Ba_3ReO_5Cl_2$ and $Sr_3ReO_5Cl_2$

| Bond | $l$ (Å) | | |
|---|---|---|---|
| | $Ba_3ReO_5Cl_2$ | $Sr_3ReO_5Cl_2$ | $Ca_3ReO_5Cl_2$ |
| Re-O1 (basal) | 1.904(4) × 4 | 1.904(4) × 4 | 1.899(5) × 2, 1.920(5) × 2 |
| Re-O2 (apical) | 1.710(10) | 1.709(8) | 1.716(8) |
| Re-Re (along $J$) | 5.7942(2) × 2 | 5.6492(4) × 2 | 5.5661(1) × 2 |
| Re-Re (along $J'$) | 6.4203(3) × 4 | 6.2434(3) × 4 | 6.3989(3) × 4 |
| Re-Re (along $J''$) | 7.5531(6) | 7.1738(6) | 5.5515(3) |

**3.2. Magnetic susceptibility.** The magnetic properties of BROC and SROC are very similar to that of CROC. As shown in Fig. 4, the temperature dependences of magnetic susceptibility have a broad peak, characteristic of low-dimensional magnets, at 11 and 29 K for BROC and SROC, respectively. There is no anomaly indicative of an LRO above 2 K for each compound. The finite magnetic susceptibility at the low-temperate limit demonstrates the absence of a spin gap.

The high-temperature susceptibilities of BROC and SROC show Curie–Weiss behaviors. The $\chi(T)$ data of BROC between 60 and 400 K and that of SROC between 100 and 400 K are fitted to the Curie–Weiss equation $\chi = (T - \Theta_W)/C + \chi_0$, where $\Theta_W$ is the Weiss temperature, $C$ is the Curie constant, and $\chi_0$ is the temperature-independent susceptibility. The fittings yield $C = 0.2741(3)$ cm$^3$ K mol$^{-1}$, $\chi_0 = -1.931(7) \times 10^{-4}$ cm$^3$ mol$^{-1}$, and $\Theta_W = -21.6(1)$ K for BROC and $C = 0.3394(3)$ cm$^3$ K mol$^{-1}$, $\chi_0 = -2.219(5) \times 10^{-4}$ cm$^3$ mol$^{-1}$, and $\Theta_W = -49.5(1)$ K for SROC. The obtained temperature-independent terms are comparable to the diamagnetic susceptibilities expected from the inner shell electrons of $-2.023 \times 10^{-4}$ cm$^3$ mol$^{-1}$ for BROC and $-1.798 \times 10^{-4}$ cm$^3$ mol$^{-1}$ for SROC. The linear temperature dependences of $(\chi - \chi_0)^{-1}$ at high-temperatures, shown in the inset of Fig. 4, confirm the validity of the fittings.

Table 2 Comparison of magnetic and thermal parameters between $A_3ReO_5Cl_2$ (A = Ba, Sr, and Ca) compounds.

| | $Ba_3ReO_5Cl_2$ | $Sr_3ReO_5Cl_2$ | $Ca_3ReO_5Cl_2$ |
|---|---|---|---|
| Weiss temperature $\Theta_W$ (K) | −21.6(1) | −49.5(1) | −37.8(1) |
| Effective moment $\mu_{eff}$ ($\mu_B$) | 1.480(3) | 1.648(1) | 1.585(2) |

| Transition temperature $T_N$ (K) | < 2 | < 2 | 1.13 |
|---|---|---|---|
| Frustration factor $f(=\Theta_W/T_N)$ | > 10.8 | > 24.8 | 33.5 |
| $J/k_B$ (K) | 19.5 | 44.9 | 40.6 |
| $J'/k_B$ (K) | 9.2 | 19.3 | 13.0 |
| Anisotropy $J'/J$ | 0.47 | 0.43 | 0.32 |
| $\alpha$ in $C$ (mJ K$^{-2}$mol$^{-1}$)* | 246.7(4) | 93.3(5) | 114.7(4) |
| $\beta$ in $C$ (mJ K$^{-4}$mol$^{-1}$)* | 6.11(3) | 2.21(2) | 1.64(1) |
| $J_d/k_B$ (K) estimated from heat capacity | 22.5(2) | 59.4(3) | 48.3(2) |

\* $C = \alpha T + \beta T^3$

The effective magnetic moments ($\mu_{eff}$) are obtained from the Curie constants to be 1.480(3)$\mu_B$ for BROC and 1.648(1)$\mu_B$ for SROC, which are smaller than 1.73$\mu_B$ expected for spin-1/2. These reductions are probably originated from the reduced $g$-factors of 1.71 for BROC and 1.90 for SROC, as is the case for CROC ($g$ = 1.78)[29]. For the 5d[1] electronic configuration, the spin and orbital angular momentums cancel each other owing to SOIs, leading to a $g$-factor reduced from 2 for a spin-only moment. The observed deviations in the $g$-factors are comparable to those observed in Cu-based quantum magnets: 2.1 < $g$ < 2.4[51,52]. Therefore, BROC and SROC are regarded as spin-1/2 quantum magnets.

The negative Weiss temperatures of $\Theta_W = -21.6(1)$ K for BROC and $\Theta_W = -49.5(1)$ K for SROC indicate dominant antiferromagnetic exchange interactions. The larger Weiss temperature of SROC indicates larger magnetic interactions than in BROC. This is consistent with the smaller distances between Re$^{6+}$ ions in SROC. The fact that more than twice difference in $\Theta_W$ is caused by only 2.6% difference in the distance, means that the magnetic interactions are sensitive to the crystal structure.

Next, the magnetic susceptibilities of BROC and SROC are fitted to the high-temperature series expansion for a spin-1/2 Heisenberg ATL antiferromagnet[21]. The [5, 5] Padé approximant[53] is used to extend the fitting range. For these fittings, we fix the $g$-factors and $\chi_0$s obtained from the Curie–Weiss fits mentioned above. As shown in Fig. 4, the theoretical model nicely reproduces each $\chi(T)$ data between the peak temperature and 400 K. A fitting between 10 and 400 K for BROC yields $J$ = 19.5(1) K and $J'$ = 9.2(1) K with their ratio of $J'/J$ = 0.47(1). For SROC, $J$ = 44.9(1) K, $J'$ = 19.3(1) K with $J'/J$ = 0.43(1) are obtained from fitting between 20 and 400 K. Compared with $J'/J$ = 0.32 in CROC[29], both SROC and BROC have larger $J'/J$ values and thus smaller anisotropies of magnetic interactions. A comparison of parameters between the three compounds is listed in Table. 2.

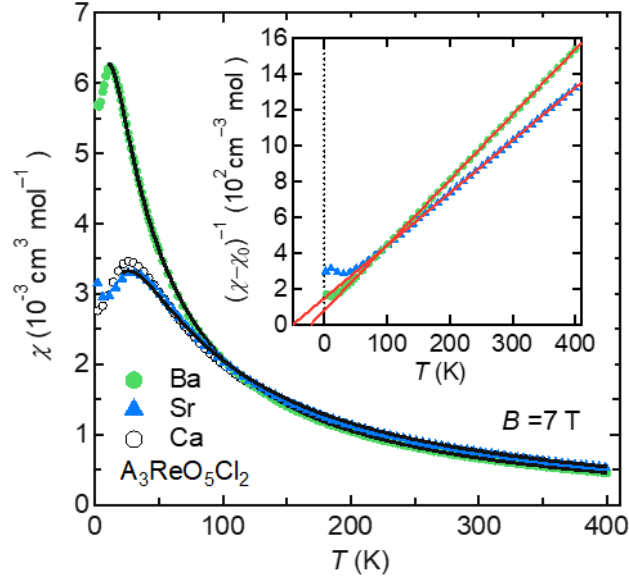

Figure 4. Temperature dependences of magnetic susceptibilities of the powder samples of BROC (green filled circle) and SROC (blue filled triangle) measured upon cooling in a magnetic field of 7 T. The data of CROC (black open circle) is also shown. The curve on each dataset represents a fit to the ATL model. The inset shows the inverses of the magnetic susceptibilities with Curie–Weiss fittings (red lines).

**3.3. Heat capacity.** Figure 5 shows the heat capacity data at low temperatures for BROC and SROC. No anomaly indicative of an LRO is observed above 2 K for each compound, which is consistent with the absence of anomaly in the magnetic susceptibilities. Considering $\Theta_W = -21.6(1)$ and $-49.5(1)$ K for BROC and SROC, respectively, LROs are strongly suppressed. The degree of suppression is estimated by the frustration factor defined as $f = \Theta_W/T_N$: $f > 10.8$ and $f > 24.8$ for BROC and SROC, respectively, suggesting strong frustration effects.

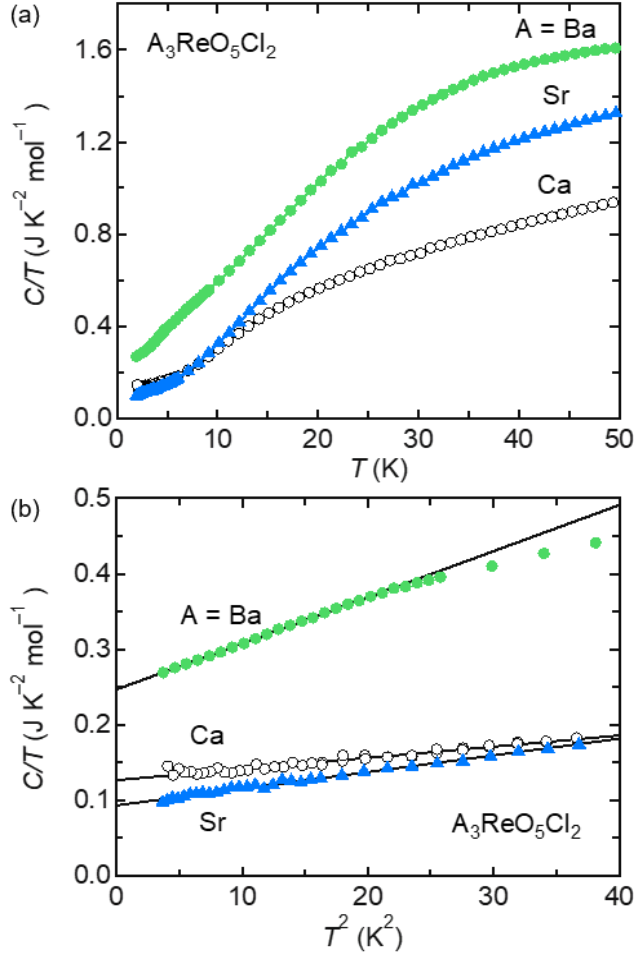

Figure 5. Heat capacity divided by temperature $C/T$ as a function of $T$ below 50 K (a) and as a function of $T^2$ below 6 K for BROC (green filled circle), SROC (blue filled triangle), and CROC (open black circle). The black lines in (b) represent fittings to the equation $C/T = \alpha + \beta T^2$.

The suppression of LROs gives rise to a large residual magnetic entropy at low temperatures. The heat capacity divided by temperature, $C/T$, shows a linear dependence against $T^2$ below 5 K with a large intercept, as shown in Fig. 5b. Fittings to the equation $C/T = \alpha + \beta T^2$, where $\alpha$ and $\beta T^2$ terms originate from spin and phonon contributions to the heat capacity, yield $\alpha = 246.7(4)$ mJ K$^{-2}$ mol$^{-1}$ and $\beta = 2.21(2)$ mJ K$^{-4}$ mol$^{-1}$ for BROC, and $\alpha = 93.3(5)$ mJ K$^{-2}$ mol$^{-1}$ and $\beta = 6.11(3)$ mJ K$^{-4}$ mol$^{-1}$ for SROC. As BROC containing heavy Ba ions has lower-energy phonon modes, the coefficient $\beta$ is larger in BROC than in SROC.

The large $T$-linear contributions to the heat capacity in BROC and SROC are common to CROC, in which $\alpha = 114.7(4)$ mJ K$^{-2}$ mol$^{-1}$ was reported[29]. The origin is considered to be a gapless-spin excitation[29], as is observed in spin-1/2 1D AF chain. In the 1D spin chain with $J$ couplings, the $T$-linear contribution is given by $C_{mag} = 2R/3J \times T$, where $R$ is the gas constant[54–56]; the larger magnetic interaction, the smaller the $T$-linear contribution. In fact, BROC with smaller $J$ has a larger $\alpha$ value than SROC. The equation gives $J_\alpha = 22.5(2)$ and $59.4(3)$ K for BROC and SROC, respectively, which are close to the corresponding $J$ values of 19.5 and 44.9 K from the ATL fits to the magnetic susceptibilities. This consistency strongly suggests that a TLL-like state is realized in BROC and SROC as a consequence of one-dimensionalization. The minor difference between $J_\alpha$ and $J$ may come from the interchain coupling $J'$. The detail of the one-dimensionalization is discussed later.

**3.4. First-principles calculations.** To understand the magnetic properties of BROC and SROC, we performed first-principles calculations. As shown in Fig. 6a, in both of the compounds, the 5d orbitals of Re and the 2p orbitals of oxygen are strongly hybridized, forming bonding states between −7.9 and −6.5 eV and antibonding states between −0.3 and 5.0 eV. Between these states, oxygen and chlorine have non-bonding states from −5.9 to −2.8 eV. The antibonding states of Re and oxygen around the Fermi energy above −0.3 eV is responsible for the magnetism. Although our density functional theory calculations predict a metallic state, Coulomb interactions $U$ should turn these compounds into Mott insulators.

Figure 6b shows the 5d orbital states around the Fermi energy that are responsible for the magnetism. The nearly dispersionless antibonding states are projected to localized Wannier orbitals, except for the $d_{x2-y2}$ orbital which is strongly hybridized with other states at higher energy, so that it cannot be mapped to a single orbital. The projected Wannier orbitals are in sequence of $d_{xy}$, degenerate $d_{xz}$ and $d_{yz}$, $d_{z2}$, and $d_{x2-y2}$ from lower energy.

The energy diagram for the crystal field splitting of the Re 5d states is schematically shown in Fig. 7a. The $e_g$ and $t_{2g}$ manifolds in the regular octahedral crystal field are separated into three nondegenerate states and one doubly degenerate state in the ReO$_5$Cl. Since the Re ion is shifted toward the apical oxygen, the $d_{xy}$ state is stabilized, while the $d_{zx}$ and $d_{yz}$ states are destabilized. Consequently, the $d_{xy}$ orbital is selected as the lowest energy state.

Single 5d electron in the Re$^{6+}$ ion, which carries spin-1/2, occupies the $d_{xy}$ orbital. In most 5d electron systems, strong SOIs as large as 0.5 eV causes entanglement between spin and orbital moments[57,58]. However, the effect of SOIs is expected to be negligible in BROC and SROC because the orbital momentum is almost quenched by the crystal field splitting of 0.8 eV between the occupied $d_{xy}$ and first excited $d_{zx}$ and $d_{yz}$ states. This is consistent with by the effective magnetic moments of $1.48\mu_B$ for BROC and $1.65\mu_B$ for SROC, which are close enough to $1.73\mu_B$ expected for pure spin-1/2. This is in sharp contrasted to single-anion 5d compounds that have unquenched orbital moments. For example, much reduced magnetic moments $0.6$–$0.7\mu_B$ are observed in the double-perovskite $5d^1$ magnets $Ba_2MgReO_6$[59] and $Ba_2NaOsO_6$[60], which is due to the partial cancelation of spin and orbital moments.

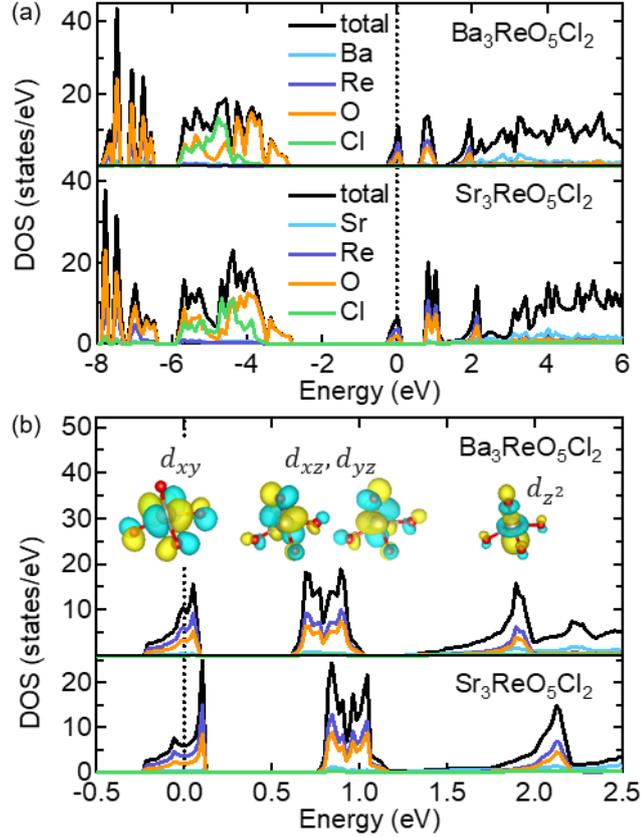

Figure 6. (a) Density of states (DOS) for BROC and SROC obtained by first-principles calculations based on the density functional theory using the program package Quantum ESPRESSO. The Fermi energy ($E_F$, vertical dotted line) is set at zero. Partial DOSs for Ba or Sr (light blue), Re (purple), O (orange), and Cl (light green) are distinguished by color lines. (b) DOS near $E_F$ and the corresponding 5d orbitals of Re mixed with the 2p orbitals

of O, the maximally localize Wannier orbitals obtained using the program package Wannier9011.

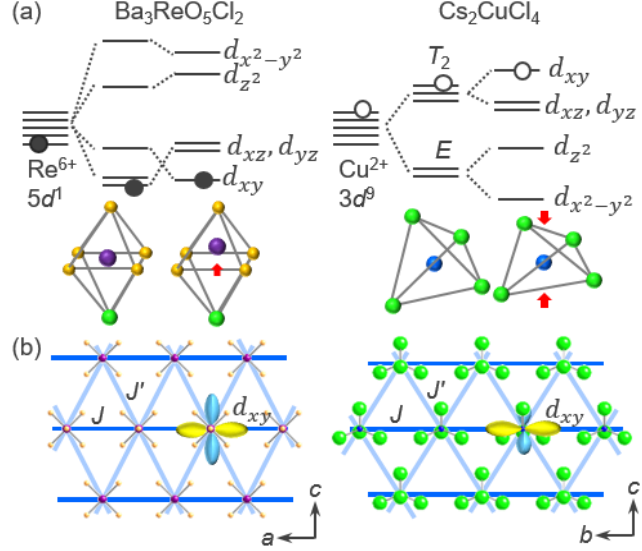

Figure 7. (a) Schematic energy diagram of d orbitals in $A_3ReO_5Cl$ and $Cs_2CuCl_4$. The lowest $d_{xy}$ orbital is singly occupied by one electron (filled circle) in $A_3ReO_5Cl$ as a result of the crystal field splitting in the distorted $ReO_5Cl$ octahedron, while the highest $d_{xy}$ orbital is singly occupied by one hole (open circle) in $Cs_2CuCl_4$ with the compressed $CuCl_4$ tetrahedron. (b) Geometric arrangements of those $d_{xy}$ orbitals in the ATLs of $A_3ReO_5Cl$ and $Cs_2CuCl_4$.

**3.5. Comparison between the three compounds.** BROC and SROC are found to be ATL antiferromagnets that exhibit one-dimensionalization by geometrical frustration, as in CROC. The anisotropies of magnetic interactions are 0.47 and 0.43, respectively, which are significantly larger than 0.32 for CROC[29].

The larger $J'/J$ values are naturally expected from the crystal structures. As shown in Fig. 7a, the yellow lobes of $d_{xy}$ orbital point to those of the neighboring $d_{xy}$ orbitals along the *a* axis in the $A_3ReO_5$ layer and overlap with each other, leading to strong interaction *J*. In contrast, the overlap of $d_{xy}$ orbitals between the adjacent chains must be smaller, giving smaller $J'$ couplings. Thus, the distance between the Re ions $l$ given in Table 1 can be an appropriate measure of magnetic interactions. The distances along $J$ are $l(J)$ = 5.7942(2), 5.6492(4), and 5.5661(1) Å[29] and along the $J'$ are $l(J')$ = 6.4203(3), 6.2434(3), and 6.3989(3) Å[29] with their ratios $l(J')/l(J)$ = 1.108, 1.105, and 1.150 for BROC, SROC, and CROC, respectively: CROC has the largest difference of 15% and BROC and SROC have similar values of 11%, which is consistent with the observed trend in $J'/J$.

These three compounds are very similar in the magnetic interactions in the 2D ATLs, while the interlayer interactions are expected to be different. As mentioned above, the Ba and Sr ions in BROC and SROC have larger coordination numbers than the Ca ions in CROC, which results in much longer Re–Re distances between the $A_3ReO_5$ layers: the Re–Re distances along the $J''$ couplings are 7.5531(6), 7.1738(6), and 5.5515(3) Å[29] for BROC, SROC, and CROC, respectively (Table 1). Therefore, one expects weaker interlayer couplings and thus a better two dimensionality for BROC and SROC than CROC.

CROC shows a 3D LRO below 1.13 K[29], which is due to the not-so-small interlayer coupling as well as additional Dzyaloshinskii–Moriya (DM) interactions; such deviations from the ideal ATL antiferromagnet sometimes induce LROs instead of the intrinsic ground states. On the other hand, we observed no LROs above 2 K for BROC and SROC. Very recent muon spin rotation experiments found no LRO down to 50 mK. We expect suppression of LRO for BROC and SROC because of the smaller interlayer couplings as mentioned above and the absence of DM interactions due to the symmetry: inversion symmetry exists at the center of the Re–Re bonds along the $J'$ couplings, and there are mirror planes perpendicular to the $a$ and $c$ axis and two-fold rotation axis along the $b$ axis exist at the center of the bonds along the $J$ couplings in BROC and SROC; no such restriction of DM interactions in CROC. Therefore, BROC and SROC are better ATL magnets compared with CROC to study the one-dimensionalization by geometrical frustration and to search for exotic phenomena.

**3.6. Comparison to $Cs_2CuCl_4$.** Here we compare $A_3ReO_5Cl_2$ (A= Ca, Sr, Ba) to $Cs_2CuCl_4$. In $Cs_2CuCl_4$, the $Cu^{2+}$ ion with the $3d^9$ electronic configuration, which carries spin-1/2, is coordinated by four chloride atoms in tetrahedral coordination. As shown in Fig. 7a, the 3d states split into triply degenerated $T_2$ manifold at higher energy and E doublet at lower energy by the crystal field. As the $CuCl_4$ tetrahedron is compressed along the $z$ axis, the $d_{xy}$ orbital in $T_2$ manifold takes the highest energy level and is occupied by one hole carrying spin-1/2[61]. As a result, the occupied $d_{xy}$ orbitals form such an ATL, as illustrated in Fig.7b. Thus, both of $A_3ReO_5Cl_2$ and $Cs_2CuCl_4$ have similar orbital arrangements with the lobes of the $d_{xy}$ orbitals extending parallel and perpendicular to the chain direction. However, the magnitudes of magnetic interactions are very different between them: small for $Cs_2CuCl_4$ ($J = 4.3$ K)[19–22], while large for $A_3ReO_5Cl$ ($J = 40.6$[29], 44.9, and 19.5 K for CROC, SROC, and BROC, respectively). This large difference probably comes from the spatially more extended nature of 5d orbitals than 3d orbitals. Thanks to the large $J$ in $A_3ReO_5Cl_2$, we can study the characteristic feature of one-dimensionalization in a wide temperature range ($T/J < 1$) in bulk properties such as magnetic susceptibilities and heat capacities.

One of two more advantages for $A_3ReO_5Cl_2$, especially BROC and SROC, compared to $Cs_2CuCl_4$, is a better two dimensionality with less interlayer couplings. The $d_{xy}$ orbitals in $Cs_2CuCl_4$ are tilted from the ATL plane, which gives rise to sizable interlayer interactions of $J''/J = 0.045(5)$[20]. In contrast, the interlayer interactions in $A_3ReO_5Cl_2$ with the $d_{xy}$ orbitals perfectly lying in the ATL plane must be much smaller ($J''/J = 0.007$ by first-principles calculations for CROC[29]); the interlayer interactions of BROC and SROC may be even smaller. The other advantage is the absence of DM interactions for BROC and SROC, as mentioned above. It is known that DM interactions are responsible for the small frustration factor of $f = 6.9$ for $Cs_2CuCl_4$[62,63]. The large frustration factors of $A_3ReO_5Cl_2$ (CROC: $f = 36$[29], SROC: $f > 24.8$, BROC: $f > 10.8$) enables us to study the QSL states. The family of compounds provides us with an ideal material platform for studying the intriguing magnetism of the ATL magnets.

## 4. CONCLUSIONS

We have successfully synthesized new spin-1/2 quantum magnets containing 5d rhenium ions, $Ba_3ReO_5Cl_2$ and $Sr_3ReO_5Cl_2$. These compounds are found to be ideal ATL quantum magnets: the crystal structure is highly two dimensional and DM interactions are absent. These features give rise to the strong suppression of LROs below 2 K, much lower than the magnetic interactions. Signatures of one-dimensionalization by geometrical frustration have been observed in the Bonner–Fisher-like magnetic susceptibility and the large $T$-linear heat capacity. Our findings demonstrate that 5d transition metal compounds provide a promising platform for the exploration of novel quantum magnets.

ACKNOWLEDGMENTS

This work was financially supported by Japan Society for the Promotion of Science (JSPS) KAKENHI Grant Number JP18K13491, JP18H04308 (J-Physics), JP19H04688 and by Core-to-Core Program (A) Advanced Research Networks.